\documentclass[%
 amsmath,amssymb,
longbibliography,
twocolumn
]{revtex4-1}

\usepackage{graphicx}% Include figure files
\usepackage{epstopdf}
\usepackage{dcolumn}% Align table columns on decimal point
\usepackage{bm}% bold math
\usepackage{physics}
\usepackage{enumitem}
\usepackage{xcolor}
\begin{document}

\preprint{APS/123-QED}

\title{A contextually objective approach\\ to the extended Wigner's friend thought experiment}
%The extended Wigner's friend thought experiment}
\author{Maxime Federico}
\affiliation{Laboratoire Interdisciplinaire Carnot de Bourgogne, CNRS - Universit\'e Bourgogne Franche-Comt\'e, UMR 6303, BP 47870, 21078 Dijon, France}
\author{Philippe Grangier}
\affiliation{Laboratoire Charles Fabry, IOGS, CNRS, Universit\'e Paris~Saclay, F91127 Palaiseau, France.}
\date{\today}

%%%% Abstract text to be placed here %%%%%%%%%%%%
\begin{abstract}
We present a discussion of the extended Wigner's friend thought experiment proposed by Frauchiger and Renner in \cite{FR}. We show by using various arguments, including textbook quantum mechanics and the ontological approach of Contexts, Systems, Modalities (CSM), that no contradiction arises if one  admits that agents must agree on what is considered as a system and what is not. In such a contextually objective approach of quantum mechanics, the apparent contradiction is automatically removed. We also discuss why this mutual agreement between agents is already implicit in the standard formulations of quantum mechanics, and why removing it leads to inconsistencies.
\end{abstract}

\maketitle

%%%%%%%%%%%%%%%%%%%%
\section{Introduction}
\label{intro}
\bigskip

Wigner's friend thought experiment has been proposed by Wigner  in 1967 \cite{wigner}. 
Recently, an extended version was proposed \cite{FR} in order to test how quantum mechanics 
describes agents who themselves use that theory to predict results of experiments, that may include other agents.
This article triggered a large number of reactions \cite{arxiv} and it is not our purpose here to analyse all of them. 
We will rather summarize the argument, and show that the claimed contradiction between three arguments is 
removed not by giving up one of these three arguments, but rather by adding a fourth one. We will justify 
this additional argument by different approaches, including e.g. textbook quantum mechanics (QM), 
and the ontological approach of Contexts, Systems, Modalities (CSM). It may be concluded that the theory presented as QM 
by Frauchiger and Renner is not the actual QM, but a different theory that is not consistent indeed. 

In Sections II and III we present the argument and the contradiction as introduced in \cite{FR},  and in Sec. IV how to remove it by adding a simple fourth requirement. Then we discuss how to relate this argument to Hardy's paradox \cite {Bub,H-paradox} (Sec. V), to textbook QM as discussed by Lalo\"e \cite{Laloe} (Sec. VI), and finally to the CSM point of view (Sec. VII). We conclude in Sec. VIII, and present some additional issues in Appendices 1 and 2. 

\section{The Frauchiger-Renner thought experiment}

 \begin{figure}[h]
\centering
\includegraphics[scale=0.35]{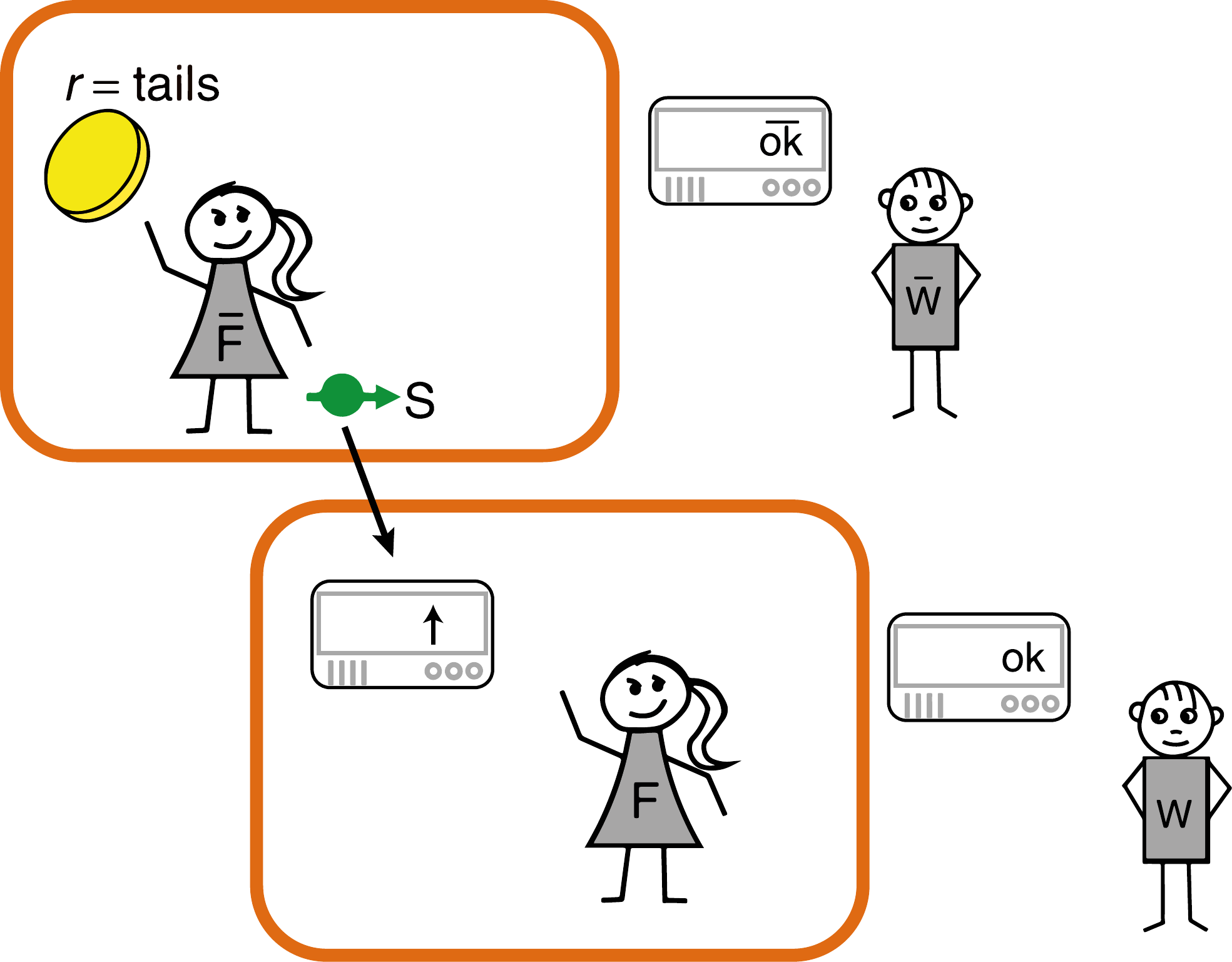}
\caption{Sketch of the thought experiment. Image adapted from \cite{FR}.}
\label{illustr exp}
\end{figure} 

The experimental protocol is the following (see Fig.~\ref{illustr exp}):
an agent $\overline{F}$ uses a quantum random generator (quantum coin) with output
$ r \in \{ heads, tails\}$. The coin quantum state is
\begin{align}
\ket{r}=\frac{1}{\sqrt{3}} \ket{h}+\sqrt{\frac{2}{3}} \ket{t},
\end{align}
where $\ket{h}$ and $\ket{t}$ are the states of $heads$ and $tails$ respectively. 
Probabilities are therefore $1/3$ for $heads$ and $2/3$ for $tails$.
During her measurement $\overline{F}$ gets entangled with the coin, 
so the state of the total system including $\overline{F}$ (and her lab) becomes
\begin{align}
\ket{\Psi}=\frac{1}{\sqrt{3}} \ket{h}\ket{\overline{F}:h}+\sqrt{\frac{2}{3}} \ket{t}\ket{\overline{F}:t},
\end{align}
using the notations \cite{Laloe} where $\ket{\overline{F}: h}$ is the state of $\overline{F}$ (and her lab)
 if she observes $heads$ and $\ket{\overline{F}: t}$ if she observes $tails$.
Depending on the result, she prepares the spin of an electron in state $\ket{\downarrow}$ 
if $r=heads$ or in state $\ket{\rightarrow}=\frac{1}{\sqrt{2}}(\ket{\downarrow}+\ket{\uparrow} )$ if $r=tails$.
Therefore the total state becomes
\begin{align}
\ket{\Psi}=\frac{1}{\sqrt{3}} \ket{h}\ket{\overline{F}:h}\ket{\downarrow}
+\sqrt{\frac{2}{3}} \ket{t}\ket{\overline{F}:t}\frac{1}{\sqrt{2}} \Big(\ket{\downarrow}+ \ket{\uparrow} \Big).
\label{spin}
\end{align}
Then, she sends this electron to a second agent $F$, located in an other lab isolated from $\overline{F}$ one (except during the exchange of the electron assumed very short) who can measure the spin projection of that electron in the basis $\{\ket{\downarrow}, \ket{\uparrow} \}$.
At this point, the total system including agents themselves is in the overall entangled state
\vspace{-2mm}
\begin{align}
\ket{\Psi}&=\frac{1}{\sqrt{3}} \Big[ \ket{h} \ket{\overline{F}: h} \ket{\downarrow} \ket{F:\ \downarrow} \nonumber \\
&+ \ket{t}\ket{\overline{F}: t} \ket{\downarrow} \ket{F:\ \downarrow} + 
	       \ket{t}\ket{\overline{F}: t} \ket{\uparrow} \ket{F:\ \uparrow} \Big], \label{state1}
\end{align} 
where  $ \ket{F:\ \uparrow}$  and $\ket{F:\ \downarrow}$ are the states of $F$ (and her lab)
observing spin $up$ or $down$.

 Two other separate agents $\overline{W}$ and $W$ can moreover perform measurements on the systems $\overline{F}$ and $F$ respectively with respect to the basis states
\begin{align}
\ket{\overline{OK}} & =\frac{1}{\sqrt{2}}\Big[ \ket{h}\ket{\overline{F}:h} 
- \ket{t}\ket{\overline{F}:t} \Big], \label{okbar}\\
\ket{\overline{fail}} & =\frac{1}{\sqrt{2}}\Big[ \ket{h}\ket{\overline{F}:h} 
+ \ket{t}\ket{\overline{F}:t} \Big],\\
\ket{OK}  & =\frac{1}{\sqrt{2}}\Big[\ket{\downarrow} \ket{ F:\ \downarrow} 
-\ket{\uparrow} \ket{ F:\ \uparrow} \Big],\\
\ket{fail}  & =\frac{1}{\sqrt{2}} \Big[\ket{\downarrow} \ket{ F:\ \downarrow}
+\ket{\uparrow} \ket{ F:\ \uparrow} \Big].
\label{ws}
\end{align}
Here we name the ``sub-experiments"
that composed the thought experiment by the name of agents corresponding to the different measurements 
$\overline{F}$, $F$, $\overline{W}$ and $W$. 
However, it should be clear that $\overline{F}$ and $F$, who were agents in the initial step, 
are then considered as systems from the point of view of $\overline{W}$ and $W$. 
Therefore the states (\ref{okbar})-(\ref{ws}) written above contain agents, labs and measured systems on which $\overline{W}$ and $W$ are supposed to make quantum measurements, so that the whole ensemble of $\overline{F}$ and $F$ agents, labs and measured systems are projected into superposition states. This is clearly not feasible in practice, but the whole idea 
is to consider this operation feasible as a thought experiment, and to examine which consequences can be drawn from it. 

%\vspace{-3mm}
Formally, considering the four situations where the active agents are either
($\overline{F}$, $F$), ($\overline{W}$, $F$), ($\overline{F}$, $W$), or ($\overline{W}$, $W$), 
by using and combining appropriate bases, $\ket{\Psi}$ can be written equivalently as
\begin{subequations}
\label{Psi}
\begin{flalign}
&\ket{\Psi}= 
\frac{1}{\sqrt{3}} \Big[ \ket{h} \ket{\overline{F}: h} \ket{\downarrow} \ket{F:\ \downarrow}  \nonumber \\
&\qquad   + \ket{t}\ket{\overline{F}: t} \ket{\downarrow} \ket{F:\ \downarrow}
	      + \ket{t}\ket{\overline{F}: t} \ket{\uparrow} \ket{F:\ \uparrow} \Big] \label{psi1}   \\
& = \sqrt{\frac{2}{3}}\ket{\overline{fail}}\ket{\downarrow}\ket{F:\ \downarrow}+
\frac{1}{\sqrt{6}}\Big[\ket{\overline{fail}} - \ket{\overline{OK}}\Big] \ket{\uparrow}\ket{F:\ \uparrow}
\label{psi2}   \\
& = \frac{1}{\sqrt{6}}\ket{h}\ket{\overline{F}: h} \Big[\ket{OK}+\ket{fail}\Big]+
\sqrt{\frac{2}{3}}\ket{t} \ket{\overline{F}:t} \ket{fail}
\label{psi3}  \\
& = \frac{1}{\sqrt{12}}\ket{\overline{OK}}\ket{OK} - \frac{1}{\sqrt{12}}\ket{\overline{OK}}
\ket{fail} \nonumber \\
& \qquad  +\frac{1}{\sqrt{12}}\ket{\overline{fail}} \ket{OK}+\frac{\sqrt{3}}{2}\ket{\overline{fail}}\ket{fail}.
\label{psi4}
\end{flalign}
\end{subequations}
Using these four expressions for the state vector $\ket{\Psi}$, one can deduce the following statements, which are close to the ones listed in \cite{FR}:
\vspace{-1mm}
\begin{itemize}[label=\textbullet]
\item $\overline{F}$ and $F$ cannot get results $heads$ and spin $up$ {\hfill [1.A]} 
\item If $\overline{W}$ gets $\overline{OK}$, then $F$ gets  spin $up$ {\hfill [1.B]} 
\item If $W$ gets $OK$, then $\overline{F}$ gets $heads$ {\hfill [1.C]} 
\item $\overline{W}$ and $W$ can get $\overline{OK}$ and $OK$ with probability $1/12$ {\hfill [1.D]} 
\end{itemize}
\vspace{-1mm}
\noindent 
If considered all true together, it is clear that these four statements lead to a contradiction, indeed from [1.D] 
 $\overline{W}$ and $W$ can get $\overline{\text{OK}}$ and OK, and in that case  $F$ and $\overline{F}$
should get spin $up$ and $heads$ from [1.B] and [1.C], contradicting [1.A]. 

However this contradiction may likely be attributed to the undefined 
status of $\overline{F}$ and $F$, who switch between being agents (able to make quantum measurements) and systems 
(being acted upon by quantum measurements). It is therefore required to clarify the definition and the role of agents. 
From here, several points of view exist and we are going to explore some of them.
\vspace{-2mm}
%%%%%%%%%%%%%%%%%%%%

\section{Frauchiger \& Renner: getting a contradiction.}
\label{FR contradiction}
\vspace{-2mm}

This point of view is based on three assumptions which allow authors of Ref \cite{FR} to show a ``no-go theorem".
\vskip 2 mm

{\bf First assumption (Q) : }
This defines what an agent must do to interpret and predict measurements.
Suppose that a system $S$ (external to the agent)  is in state $\ket{\psi}$ of a Hilbert space at time $t_0$; 
an outcome $x$ can be measured on $S$ at time $t$ with respect to a family of projectors $\{\pi_x\}$ in Heisenberg representation.
If $\bra{\psi}\pi{_\xi} \ket{\psi}=1$, then the agent can conclude that $x=\xi$ at time $t$. 
From the point of view on this individual agent, this theory looks like quantum mechanics, and it will be used 
successively from the ``subjective" point of view of different agents, in order to get a serie of statements.

First, if $r=heads$, the spin is in state $\ket{\downarrow}$, and a measurement made by agent $F$ with respect to the basis
$\{\pi_{\downarrow}=\ket{\downarrow}\bra{\downarrow}, \pi_{\uparrow}=\ket{\uparrow}\bra{\uparrow} \}$ gives
%\begin{align}
$\bra{\downarrow}\pi_{\downarrow}\ket{\downarrow}=1.$
%\end{align}

Therefore agents $\overline{F}$  and $F$ can conclude that if $\overline{F}$ gets $heads$, $F$ will get spin $down$; 
which is logically equivalent to: if $F$ gets spin $up$, $\overline{F}$ got $tails$, in agreement with eq. (\ref{psi1}). 
By considering successively the different pairs of agents, other statements can be obtained and lead to
\begin{itemize}[label=\textbullet]
\item If $F$ gets spin $up$, then $\overline{F}$ gets $tails$ {\hfill [2.A]} 
\item If $\overline{W}$ gets $\overline{OK}$, then $F$ gets spin $up$ {\hfill [2.B]}
\item If $\overline{F}$ gets $tails$, then $W$  gets $fail$ {\hfill [2.C]}
\item $\overline{W}$ and $W$ can have $\overline{\text{OK}}$ and OK {\hfill [2.D]}
\end{itemize}
\noindent
which are either the same or logically equivalent statements compared to Section \ref{intro} (in the same order). 
\\

{\bf Additional assumptions (C) and (S):}
These two hypotheses stipulate that agents can trust predictions made by other agents, irrespective of their status, and that 
all predictions apply in the same universe, avoiding Everett's multi world interpretation. It is under those three assumptions that a contradiction arises. 

Indeed, if statement [2.D] is valid, [2.B] implies that $F$ gets spin $up$, [2.A] that $\overline{F}$ gets $tails$, and [2.C] that $W$ gets $fail$. The last statement clearly contradicts the starting point that $W$ gets $OK$.  
Using their assumptions as just shown, the authors of \cite{FR} can thus formulate the following  no-go theorem:
\begin{align}
\text{(Q)} + \text{(C)} + \text{(S)} \Rightarrow   \text{Contradiction}
\end{align}
which has been formulated as the statement ``quantum theory cannot consistently describe the use of itself" \cite{FR}. 

%%%%%%%%%%%%%%%%%%%%

\section{Removing the contradiction by an agreement between agents.}
\label{assump a}
\vskip  2mm

A major remark is that, though it is presented as equivalent to quantum mechanics (QM), 
the assumption (Q) of Frauchiger and Renner implies a very unusual use of the quantum formalism, 
where ``self-proclaimed" agents may become systems for other agents.  
It means that the usual measurement postulate can hardly be applied in a consistent way \cite{Laloe}.  
A simple way to remove this problem, and also the contradiction, is to consider the additional assumption 
\\

\noindent (A) {\it  All agents must agree on the definition of the quantum system to which they apply assumption (Q); 
as a consequence, no agent should be included in what another agent considers as the measured system. }
\\

\noindent  The second part of (A), i.e. that no agent should be included in what another agent considers as the measured system, 
is a joint consequence of (Q), stating that the agents apply QM to a system external to themselves, and of the first part of (A), 
i.e. the agents agreement on the definition of the system. Then the quantum system is the same for all agents, and contains no agent. 

Given (A), it is clear that the previous contradiction vanishes since the four statements cannot be true together: 

(i)  if $\overline{F}$ and $F$ are agents, the first statement is true, but then the other three make no sense since $\overline{F}$ and $F$  cannot behave as (superposable) systems from the point of view of $\overline{W}$ and $W$;

(ii)  if $\overline{F}$ and $F$  are systems, then the four lines correspond to the agents $\overline{W}$ and $W$ 
making four incompatible measurements on the global system composed of $\overline{F}$, $F$, the coin and the spin. 
Incompatible measurements cannot be true together as it is usual in QM, such that there is no contradiction. 

Therefore, with all of the four assumptions, we obtain 
\begin{align}
\text{(Q)} + \text{(C)} + \text{(S)} +  \text{(A)}  \Rightarrow   \text{No contradiction}
\end{align}
which  means that assumption  (A) restricts (Q) to some safe range where no contradiction  arises.  
This could be summarized as : ``Objective quantum mechanics can describe the use of itself'', 
where ``objective" is  meant as requiring a mutual agreement between all the agents. 

In the following we will develop these arguments, and in particular give more justifications and illustrations of the points (i) and (ii) above, 
as well as some intermediary situations. We will also see that assumption (A) can take different forms depending on which point of view we adopt, but always removes the contradiction.

%%%%%%%%%%%%%%%%%%%%

\section{Looking at $\overline{W}$ and $W$ as agents:  link to Hardy's paradox}
\label{hardyparadox}
\vskip  2mm

The Frauchiger and Renner (F\&R) paradox is, as it was pointed out by Bub \cite{Bub}, not only a new formulation of Wigner's friend
experiment but also a twisted version of Hardy's paradox originally presented to show a similar result as Bell's theorem \cite{Bell}. To do that, Hardy \cite{H-paradox} used a Mach-Zehnder interferometer for electrons and positrons in order to get four entangled states which are similar to those of the F\&R Gedankenexperiment. The difference arises only with the way one imposes that these four statements must be true together. In the following, we will assume that a hidden variable exists to describe the various states of eq. (\ref{Psi}), in agreement with the formulation proposed in \cite{H-paradox}.

Let $\lambda$ be the hidden variable which describes the states before measurements and assume that QM is a local realist theory.
It means that results of measurements do not depend on the choice of measurement done by other agents. We introduce notations $\overline{F}(h,\lambda)=1$ if $\overline{F}$ measures the state $\ket{h}$, and $\overline{F}(h,\lambda)=0$ otherwise. The same notations will be applied to other agents and their respective measurable states. The statements obtained in the preceding sections can be recovered using the expressions of $\ket{\Psi}$ in the different bases. Indeed, from (\ref{psi1}) and for every experiment described by a given value of $\lambda$,
\begin{equation}
\overline{F}(h,\lambda)F(\uparrow,\lambda)=0.
\end{equation}
From (\ref{psi2}) we additionally have
\begin{equation}
\text{if} \ \overline{W}(\overline{OK},\lambda)=1 \ \text{then} \ F(\uparrow,\lambda)=1,
\end{equation}
from (\ref{psi3})
\begin{equation}
\text{if}\ W(OK,\lambda)=1 \ \text{then}\ \overline{F}(h,\lambda)=1,
\end{equation}
and from (\ref{psi4})
\begin{equation}
\overline{W}(\overline{OK},\lambda)W(OK,\lambda)=1 \ \text{for $1/12$th of experiments},
\end{equation}
which can be summed up in the four following statements completely equivalent to [1.A]-[1.D]:
\vskip 5mm

\begin{itemize}[label=\textbullet]
\item $\overline{F}(h,\lambda)$ and $F(\uparrow,\lambda)$ cannot occur at the same time
\item $ \overline{W}(\overline{OK},\lambda)$ is true $\Rightarrow F(\uparrow,\lambda)$  is true
\item $W(OK,\lambda)$ is true $\Rightarrow \overline{F}(h,\lambda)$  is true
\item $\overline{W}(\overline{OK},\lambda)$ and $W(OK,\lambda)$ occur one time out of $12$
\end{itemize}
One can see that even if the assumptions and consequently the formalism are not exactly the same, the contradiction arises in the same way as when the four states of eq. (\ref{Psi}) are considered simultaneously true. In that case inspired by Hardy's paradox, this is due to the hidden variable hypothesis. Therefore, it appears that in both situation, the contradiction is not due to QM itself but to the assumption that all conclusions drawn from eq. (\ref{Psi})  are simultaneously true. This explains why the use of assumption (A) removes the contradiction, by making clear that they are incompatible in a quantum sense.  

%%%%%%%%%%%%%%%%%%%%

\section{Textbook QM: projective measurements define agents}
\label{projective measurement}
\vskip  2mm

Another approach to the F\&R contradiction can be found in \cite{Laloe}. This point of view stipulates that defining agents
is equivalent to define where the projective measurements are done, i.e. where the state is projected. 
For instance, without any projection, the state of the total system after ``measurements" by $\overline{F}$, $F$, $\overline{W}$ and $W$ should be written as
\begin{align}
&&\ket{\Psi}= \frac{1}{\sqrt{12}}\ket{\overline{OK}}\ket{\overline{W}:\overline{OK}} \ket{OK}\ket{W:OK}
\nonumber \\
&&-\frac{1}{\sqrt{12}}\ket{\overline{OK}}\ket{\overline{W}:\overline{OK}}\ket{fail}\ket{W:fail}\nonumber 
\nonumber \\
&& +\frac{1}{\sqrt{12}}\ket{\overline{fail}}\ket{\overline{W}:\overline{fail}} \ket{OK}\ket{W:OK}
\nonumber \\
&&+\frac{\sqrt{3}}{2}\ket{\overline{fail}}\ket{\overline{W}:\overline{fail}}\ket{fail}\ket{W:fail}.
\end{align}
In this state, all observers are entangled as if we had considered Everett's multi-world interpretation. Then all possibilities can be investigated by choosing where the projections are made. The calculations are detailed in \cite{Laloe} and Appendix 2, here we simply describe qualitatively the results.

First, we consider $\overline{F}$ and $F$ as agent (point (i) of Section
\ref{assump a}), and project states in agreement with their measurements. 
The result (see Appendix 2) is that whatever $\overline{W}$ and $W$ measure, they cannot obtain informations on spin state or coin state. Therefore, the statements [2.B] and [2.C] of Section \ref{FR contradiction} cannot be formulated and the contradiction does not arise. 

In the second option which considers $\overline{F}$ and $F$ as systems, and only $\overline{W}$ and $W$ as agents, $\overline{W}$ cannot say anything about the spin state, therefore using this way of defining agents and writing states after measurements, also prevent from any contradiction. 

It is also instructive to  consider the slightly modified situation where $\overline{F}$ can ``protect" her result  by using
an other qubit stored in her lab, in such a way that it escapes to $\overline{W}$'s measurement (refered as ``hidden qubit" \cite{Laloe}).
If $\overline{F}$ observes ``heads", the hidden qubit is in state $\ket{\overline{G}:h}$ and in state $\ket{\overline{G}:t}$ 
if she observes ``tails". The overall state can thus be written as 
\begin{align}
\ket{\Psi}&=\frac{1}{\sqrt{3}} \Big[ \ket{h} \ket{\overline{F}: h}\ket{\overline{G}:h} \ket{\downarrow} \ket{F:\ \downarrow}
\nonumber \\  
&\quad + \ket{t}\ket{\overline{F}: t}\ket{\overline{G}:t} \ket{\downarrow} \ket{F:\ \downarrow} \nonumber \\  
& \quad+ \ket{t}\ket{\overline{F}: t}\ket{\overline{G}:t} \ket{\uparrow} \ket{F:\ \uparrow} \Big]. \label{hidd}
\end{align} 
In the following, to simplify notations, we will replace $\ket{h} \ket{\overline{F}: h}$ by $\ket{h}$, $\ket{t} \ket{\overline{F}: t}$ by $\ket{t}$, $ \ket{\downarrow} \ket{F:\ \downarrow}$ by $\ket{\downarrow}$ and $\ket{\uparrow} \ket{F:\ \uparrow}$ by $\ket{\uparrow}$ since agents (and their labs) are always in the state corresponding to the outcome they have measured. We also condense the hidden qubit notation in $\ket{h_G}$ and $\ket{t_G}$. The state becomes
\begin{align}
\ket{\Psi}=\frac{1}{\sqrt{3}}\Big[\ket{h}\ket{h_G}+\ket{t}\ket{t_G}\Big]\ket{\downarrow}+\frac{1}{\sqrt{3}}\ket{t}\ket{t_G}\ket{\uparrow},
\end{align}
and the bases used by $\overline{W}$ and $W$ to perform their respective measurements are
\begin{align}
\ket{\overline{OK}}=\frac{1}{\sqrt{2}}(\ket{h}-\ket{t}),  \; \;
\ket{\overline{fail}}=\frac{1}{\sqrt{2}}(\ket{h}+\ket{t}),\\
\ket{OK}=\frac{1}{\sqrt{2}}(\ket{\downarrow}-\ket{\uparrow}),  \; \;
\ket{fail}=\frac{1}{\sqrt{2}}(\ket{\downarrow}+\ket{\uparrow}).
\end{align}
The global state can then be put in the form
\begin{align}
\ket{\Psi} & = \ket{\overline{OK}}\ket{OK} \frac{1}{\sqrt{12}}\ket{h_G} \nonumber \\
&\quad + \ket{\overline{OK}}\ket{fail} \Big[ \frac{1}{\sqrt{12}}\ket{h_G} - \frac{1}{\sqrt{3}}\ket{t_G} \Big]  \nonumber \\
&\quad + \ket{\overline{fail}}\ket{OK}\frac{1}{\sqrt{12}}\ket{h_G} \nonumber \\
&\quad + \ket{\overline{fail}}\ket{fail} \Big[ \frac{1}{\sqrt{12}}\ket{h_G} + \frac{1}{\sqrt{3}}\ket{t_G} \Big]. \label{hq}
\end{align}
As it was pointed out in \cite{Laloe}, the state $\ket{t_G}$ is only correlated with the result $\ket{fail}$ which is not surprising since when $\overline{F}$ sends a spin in the state $\ket{\rightarrow}$, it is orthogonal to $\ket{OK}$. It means that the hidden qubit keeps a memory of $\overline{F}$'s measurement and therefore all statements derived before do not hold anymore. We emphasize that there is no need to read out the value of the hidden qubit: as in an interference experiment, it is enough that the ``which path information" is stored somewhere to forbid the quantum superposition. 
\\
 
It is also interesting to investigate different scenarii for the two accessible states of the hidden qubit e.g. if $\bra{h_G}\ket{t_G}=1$ i.e. $\ket{h_G}$ and $\ket{t_G}$ are the same state, denoted for instance $\ket{G}$, then the overall system is in the following state
\begin{align}
\ket{\Psi} & =\frac{1}{\sqrt{12}} \ket{\overline{OK}}\ket{OK} \ket{G}
- \frac{1}{\sqrt{12}} \ket{\overline{OK}}\ket{fail} \ket{G}  \nonumber \\
 &\quad+ \frac{1}{\sqrt{12}} \ket{\overline{fail}}\ket{OK}\ket{G} 
 + \frac{\sqrt{3}}{2} \ket{\overline{fail}}\ket{fail}\ket{G},
\end{align}
which is the same as state (\ref{psi4}) without the hidden qubit. $\ket{G}$ can be factored out and the hidden qubit plays no role since it is no more entangled with $\overline{F}$'s results. Otherwise, if $\bra{h_G}\ket{t_G}=0$ i.e. the qubit states are orthogonal and the overall state is identical to (\ref{hq}). If we project this global state onto each state of the hidden qubit
\begin{align}
\bra{h_G}\ket{\Psi}&=\frac{1}{\sqrt{12}}\Big[ \ket{\overline{OK}}\ket{OK} + \ket{\overline{OK}}\ket{fail} 
\nonumber \\  &\quad + \ket{\overline{fail}}  \ket{OK} 
+ \ket{\overline{fail}}\ket{fail} \Big]   \nonumber \\
&=\frac{1}{\sqrt{3}} \, | h \rangle | \overline F : h \rangle \frac{\ket{OK} +\ket{fail}}{\sqrt{2}} ,  \\
\bra{t_G}\ket{\Psi} &= \frac{1}{\sqrt{3}} \Big[ \ket{\overline{fail}}\ket{fail}-\ket{\overline{OK}}\ket{fail}\Big]  \nonumber \\
&=  \sqrt{\frac{2}{3}} \,  \ket{t}\ket{ \overline F : t }       \ket{fail},
\end{align} 
we recover the situation where $\overline{F}$ is an agent, and $\overline{W}$ cannot project her in a superposition. 
\\

This calculation has two consequences: first, it gives a way to make a transition between the different options, by allowing $\bra{h_G}\ket{t_G}$ to take any value between 0 and 1. 
We emphasize that this calculation makes sense in a framework where the only agents are $\overline{W}$ and $W$. 
If  $\bra{h_G}\ket{t_G}=0$ then $\overline{F}$  and $F$ should be considered as decohered systems, as they appear in the usual theory of 
decoherence after tracing out the ``ancilla" qubits. Second, it provides a ``toy model" for a situation where a system actually does not behave as such: if $\overline{W}$ tries to make a projective 
measurement on $\overline{F}$, including her lab and all the surrounding environment, it is enough that a single qubit escapes from the action 
of $\overline{W}$ to make the projective measurement failed. When $\overline{F}$ is a macroscopic system, it can be considered impossible for 
$\overline{W}$ to get full control on every single qubit entangled with $\overline{F}$; then considering $\overline{F}$ as a system makes no sense. 
\\

Following the idea of Section \ref{assump a}, this can be seen  as a reformulation of assumption (A):

\noindent (A) {\it  All agents must agree on the definition of the quantum system to which they apply assumption (Q), that is to say where
the state projection is done; as a consequence, no agent should be included in what another agent considers as the measured system.}

\bigskip

%%%%%%%%%%%%%%%%%%%%

\section{Contexts, Systems and Modalities: the  CSM approach. }
\label{CSM}

\begin{figure}[h]
\centering
\includegraphics[width = 0.51 \columnwidth]{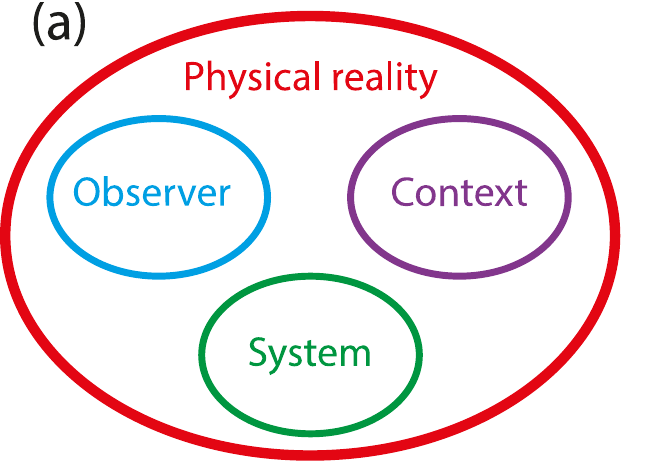}
\includegraphics[width = 0.52 \columnwidth]{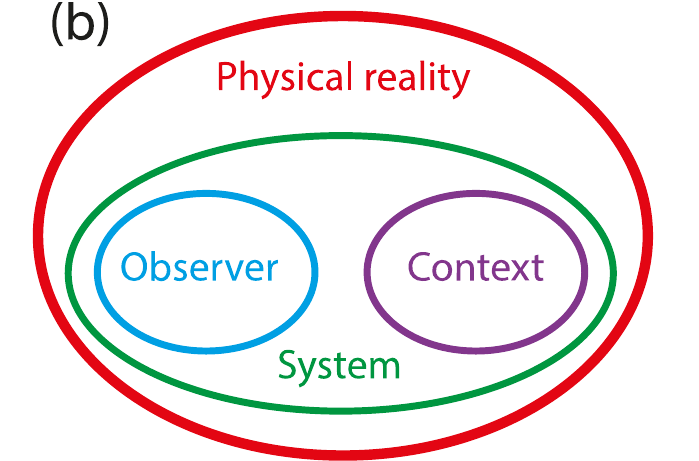}
\includegraphics[width = 0.46 \columnwidth]{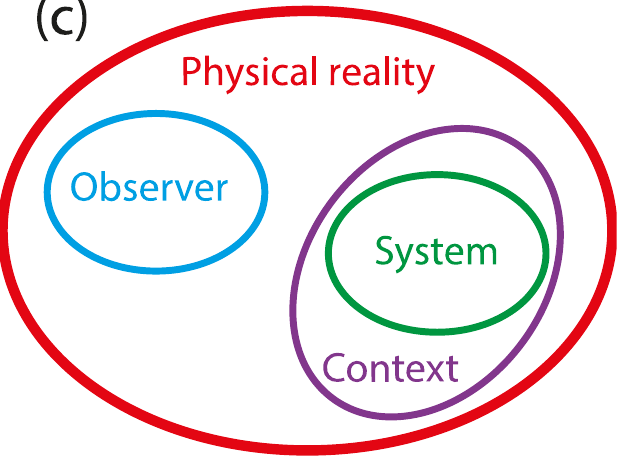}
\caption{Graphic representations of different ontologies:
(a)~Classical ontology: the observer can know the ``real'' physical properties of the system, and the context is only used as an auxiliary 
tool for measurements. (b) Usual quantum ontology: through successive ``entangling'' interactions and unitary evolution, the system will 
include the context, and also (ultimately) the observer or ``agent", whose status may be problematic. (c) CSM ontology: the context appears always between the system and the 
observer, and definite values of the relevant physical properties (modalities) are attributed jointly to the system and the context.}
\label{ont}
\end{figure}

Here, we want to show that in the framework of the CSM (Contexts, Systems and Modalities) approach, developped by A. Auff\`eves and P. Grangier \cite{pg,CSM1,CSM2,qmyst}, the assumption (A) introduced and discussed above is no more an assumption but a theorem. We recall some basics of CSM which starts with a new physical ontology (i.e. a definition of all objects that are described by the theory), by considering  three entities:
\begin{itemize}[label=\textbullet]
\item System = subpart of the world, isolated well enough to have physical properties that can be studied.
\item Context = all other physical systems external and possibly in contact with the studied one (ex: measurement apparatus, labs, etc); in CSM, the context corresponds to a classically defined measurement setup, and it acts like an interface between the system and the observer, see Fig. \ref{ont}. 
\item Modalities = the set of answers (or values)  that can be predicted with certainty and obtained repeatedly for a given system in a given  context.
\end{itemize}
These definitions are bound together by the following rule (or axiom): in quantum mechanics, modalities are attributed jointly to the system and the  context. It means that the quantum properties of a system do not ``belong" to the system alone, but to the system within a context. 

In order to give an operational content to CSM, we need a second axiom called Quantization Principle:

(i) For each well-defined system and context, there is a discrete number N of mutually exclusive modalities. 
This number N depends on the system, but does not depend on any particular context.

(ii) Modalities, when defined in different contexts, are generally not mutually exclusive, and they are said to be ``incompatible".
\vskip 4 mm

CSM is based on a physically realistic point of view,  i.e. physics is applied to objects which exist independently from any observer.
However, the ``object" for CSM is a system within a context, which makes it highly non-classical. 
As it can be shown on Figure \ref{ont}(c),  in CSM the observer is outside of the context, illustrating again that the (objective) state of the context must be taken into account in defining the modality. Such a 
``contextual objectivity" \cite{pg} is quite different from the classical ``absolute objectivity" of Figure \ref{ont}(a), 
but it fully agrees with quantum reality as deduced from empirical evidence.
\vskip 4 mm

\begin{figure}[h]
\centering
\includegraphics[width = 0.51 \columnwidth]{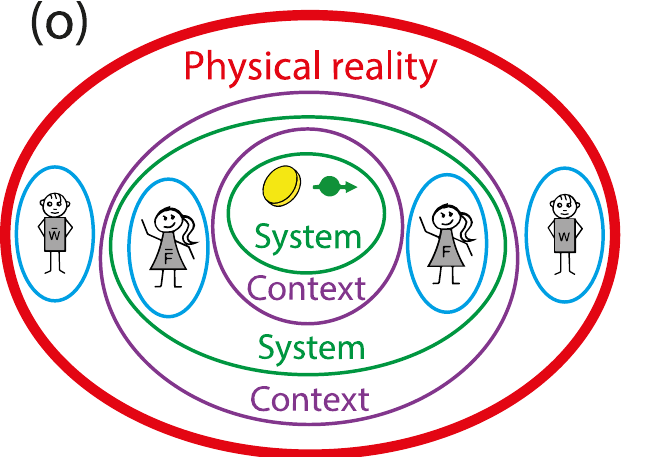}
\includegraphics[width = 0.51 \columnwidth]{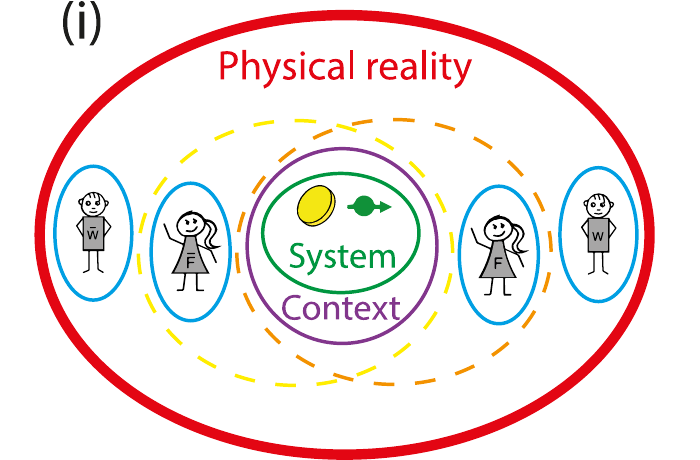}
\includegraphics[width = 0.46 \columnwidth]{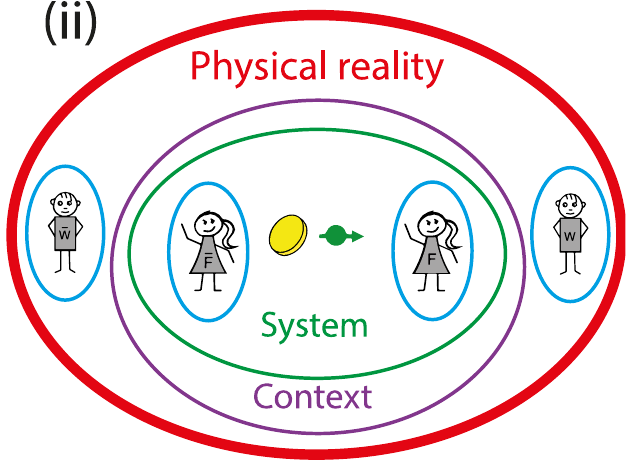}
\caption{(o) Naive way to see Wigner's friends thought experiment. Such a picture is  impossible for CSM, because a (quantum) system cannot  include a (classical)  context. (i) Case where $\overline{F}$ and $F$ are agents making quantum measurements on systems with their ``own" classical contexts.  
(ii) Second option where $\overline{F}$ and $F$ are considered as systems.}
\label{solution}
\end{figure}

Now, coming back to the original problem, CSM was hidden so far but in fact already there. In order to define systems and contexts according to CSM,  we have to avoid mixing notations  between agents, labs and experimental set ups. First, as it was 
pointed out before, there is only one ultimate reality which imposes that a context cannot be included in a system and vice versa (see Figure \ref{solution} (o)).  For consistency, different agents must agree on the definition of systems and contexts which is exactly the purpose of assumption (A), formulated according to CSM as: 
\\

\noindent (A) {\it  All agents must agree on the definition of the system and context to which they apply assumption (Q); 
as a consequence, no agent should be included in what another agent considers as the measured system. }
\\

Again, several cases are possible, corresponding e.g. to point (i) and (ii) of Section \ref{assump a}, and are drawn in 
Figure \ref{solution}. In case (i), $\overline{F}$ and $F$ perform quantum measurements on their systems within their respective contexts. Then $\overline{W}$ and $W$ cannot make quantum measurements on them (composite systems in yellow and orange dash lines of Figure \ref{solution}), and the F\&R reasoning does not hold. In case (ii), $\overline{F}$ and $F$ are considered as systems so $\overline{W}$ and $W$ can perform quantum measurements on those systems in two different contexts on each side (i.e. projecting $\overline{F}$ or $F$ in superposed or non-superposed states). The modalities defined in these different contexts are incompatible and thus not simultaneously true, no contradiction can arise. 

To conclude, we have shown that within the CSM interpretation, the F\&R contradiction is removed without additional assumption because CSM is already an ``contextually objective" theory \cite{pg}, where agents cannot become systems for other agents.

\section{Conclusion}
\label{conclusion}
\vskip 2mm

To conclude, we have presented several ways to escape the ``no-go theorem" introduced by F\&R. Our arguments can be summed up by {\bf ``Contextually objective quantum mechanics can describe the use of itself". }

This statement does not allow every single world interpretation of quantum mechanics to be self-consistent, but only those which rely on some form of objectivity. 
A consequence is that the (single world) quantum rules must introduce a clear distinction between agents and systems, embedded in the additional assumption (A). 
Then we have shown that in the framework of CSM approach,  no F\&R contradiction can arise since CSM already contains (A) as a theorem, deduced from the 
central concept of contextual objectivity \cite{pg} which governs quantum measurements in the CSM framework. 

%%%%%%%%%%%%%%%%%%%%

\section*{Appendix 1: About the time evolution of the experiment}

In their article \cite{FR}, F\&R insist on the temporal order of measurements made by successive agents,  associated to changing the
status of $\overline{F}$ and $F$ between system and context. In the CSM interpretation, changing the 
agents status in a same experiment cannot occur, therefore the chronology is not relevant. 
It could be if we had adopted Everett's point of view also 
called ``multi-world"; but this interpretation is in conflict with assumption
(S), so the F\&R no-go theorem also does not apply. 
\\

More precisely, through the usual quantum or CSM point of view, an ideal measurement takes place as follows: 
(a) entanglement of the system with the pointer (also quantum) and then (b) reading of the pointer, which brings the result at
the single classical context level.
From the Everett's point of view, there is no single classical context level but many possible results which are associated to 
differents universe branches. Then it is, in principle, possible to reverse the evolution of these branches, 
rewriting  the history as if no measurement were done, 
and to make another different measurement, creating again new branches. 
So in Everett's point of view, temporal order makes sense to describe first a measurement by $\overline F$, 
and then its ``erasure" by $\overline W$, followed by a new measurement in the $(\overline{OK};  \overline{fail}$) basis.

In CSM, the measurement can be reversed only if it has not reached the context level, i.e.  between steps (a) and (b). 
When the measurement is over (step b), the result is macroscopic and unique 
and the modalities corresponding to other results in other contexts cannot coexist.
Explicitly, if $\overline{F}$ and $F$ are superposable systems, we can measure them in superposed bases
\begin{align}
\overline{\cal S}=\{\overline{OK};\overline{fail}\}, \qquad
{\cal S}=\{OK; fail\} \nonumber
\end{align}
or in non superposed ones 
\begin{align}
\overline{\cal N}=\{heads; tails\}, \qquad
{\cal N}=\{ up; down\}. \nonumber
\end{align}
The four situations we have discussed correspond to four combined measurements $(\overline{\cal N}, {\cal N})$, $( \overline{\cal N}, {\cal S})$, $(\overline{\cal S}, {\cal N})$, $(\overline{\cal S}, {\cal S})$.
In that case, temporal order is not relevant since only one of these four measurements can be made, as in Hardy's paradox, such that there is a unique objective macroscopic world in which QM is consistent.

We note also that the CSM point of view makes a clear distinction between the usual ``pre-measurement" stage, that is entangling the system with a probe in a reversible way, and the actual irreversible measurement that brings the result to the context level. It can be told that the F\&R paradox fails because it mixes these two stages in an inappropriate way, as explained in details in \cite{zukov}.

%%%%%

\section*{Appendix 2: Adding a hidden qubit}
\label{hidden qubit}
\vskip  2mm

Another approach to the F\&R contradiction can be found in \cite{Laloe}. This point of view stipulates that defining agents
is equivalent to define where projective measurements are done, that is to say, where mathematically the state is projected. 
For instance, without any projection, the state of the total system after ``measurements" by $\overline{F}$, $F$, $\overline{W}$ and $W$ is
\begin{align}
\ket{\Psi}&= \frac{1}{\sqrt{12}}\ket{\overline{OK}}\ket{\overline{W}:\overline{OK}} \ket{OK}\ket{W:OK} \nonumber \\
&\quad-\frac{1}{\sqrt{12}}\ket{\overline{OK}}\ket{\overline{W}:\overline{OK}}\ket{fail}\ket{W:fail}\nonumber \\
&\quad +\frac{1}{\sqrt{12}}\ket{\overline{fail}}\ket{\overline{W}:\overline{fail}} \ket{OK}\ket{W:OK} \nonumber \\
&\quad+\frac{\sqrt{3}}{2}\ket{\overline{fail}}\ket{\overline{W}:\overline{fail}}\ket{fail}\ket{W:fail}.
\end{align}
In this state, all observers are entangled as if we had considered Everett's multi-world interpretation.
\\

Then all possibilities can be investigated for the projections. First, we consider $\overline{F}$ and $F$ as agent (point (i) of Section
\ref{assump a}). Therefore, they have to project states in agreement with their measurements. According to $\overline{F}$, two states 
are available
\begin{align}
\ket{\Psi}_{tails}& =\ket{t}\ket{\overline{F}: t}\frac{1}{\sqrt{2}} \big(\ket{\downarrow}+\ket{\uparrow}\big),  \\
\ket{\Psi}_{heads}& =\ket{h}\ket{\overline{F}: h}\ket{\downarrow}.
\end{align}
If then we consider $F$'s projections which lead to many cases depending of the result she gets but also of $\overline{F}$'s results, we have the states
\begin{align}
\ket{\Psi_\downarrow}_{tails}&=\ket{t}\ket{\overline{F}: t}\ket{\downarrow}\ket{F:\ \downarrow}, \\
\ket{\Psi_\uparrow}_{tails}&=\ket{t}\ket{\overline{F}: t}\ket{\uparrow}\ket{F:\ \uparrow},  \\
\ket{\Psi_\downarrow}_{heads}&=\ket{h}\ket{\overline{F}: h}\ket{\downarrow}\ket{F:\ \downarrow},
\end{align}
or equivalently written in the other bases
\begin{align}
\ket{\Psi_\downarrow}_{tails}&=\frac{1}{2}\Big[\ket{\overline{fail}}-\ket{\overline{OK}}\Big] \Big[ \ket{fail}+\ket{OK} \Big], \\
\ket{\Psi_\uparrow}_{tails}&=\frac{1}{2}\Big[\ket{\overline{fail}}-\ket{\overline{OK}}\Big] \Big[ \ket{fail}-\ket{OK} \Big], \\
\ket{\Psi_\downarrow}_{heads}&= \frac{1}{2}\Big[\ket{\overline{fail}}+\ket{\overline{OK}}\Big] \Big[ \ket{fail}-\ket{OK} \Big].
\end{align}
Here we see that all these states are products; whatever $\overline{W}$ and $W$ measure, they cannot obtain informations on spin 
state or coin state. Therefore, statements [2.B] and [2.C] of Section \ref{FR contradiction} are wrong and no contradiction can
appear.
%\bigskip

The second option is to consider $\overline{F}$ and $F$ as systems, and %only 
$\overline{W}$ and $W$ as agents (point (ii) of Section 
\ref{assump a}). Before they perform measurements, the system is in state (\ref{state1}). Then, two outcomes are possible for 
$\overline{W}$ and give the states
\begin{align}
\ket{\Psi}_{\overline{OK}} 
&=-\ket{\uparrow}\ket{F:\ \uparrow} \ket{\overline{OK}} \ket{\overline{W}: \overline{OK}} \label{OKbar + spin} \nonumber  \\
& = \frac{1}{\sqrt{2}}\Big[\ket{OK}-\ket{fail} \Big] \ket{\overline{OK}} \ket{\overline{W}: \overline{OK}},   \\
 %\label{Wbar about W}  \\
\ket{\Psi}_{\overline{fail}}  &=\Big[2\ket{\downarrow}\ket{F:\ \downarrow}+\ket{\uparrow}\ket{F:\ \uparrow}\Big]\ket{\overline{fail}}\ket{\overline{W}: \overline{fail}}
%\label{failbar + spin} 
\nonumber \\
& = \Big[ \frac{3}{\sqrt{2}}\ket{fail}\ket{W: fail} + \frac{1}{\sqrt{2}}\ket{OK}\ket{W: OK} \Big] \nonumber \\ 
& \quad\otimes \ket{\overline{fail}}  \ket{\overline{W}: \overline{fail}}. 
%\label{Wbar and W}
\end{align}
%
%\end{document}
We see again, that the result $\overline{OK}$ is only correlated to spin $up$, 
%(\ref{OKbar + spin}), 
and that $\overline{W}$ cannot predict anything on $W$'s measurement since both results $OK$ and $fail$ are still available. %(\ref{Wbar about W}).  
In the second option where $\overline W$ gets $\overline{fail}$, no predictions can be obtained about the spin state. 
%(\ref{failbar + spin}). 
If we furthemore add the projection made by $W$ we obtain
\begin{align}
\ket{\Psi_{OK}}_{\overline{OK}} & = \ket{OK}\ket{W: OK}\ket{\overline{OK}} \ket{\overline{W}: \overline{OK}}, \\
\ket{\Psi_{fail}}_{\overline{OK}} & =-\ket{fail}\ket{W: fail}\ket{\overline{OK}} \ket{\overline{W}: \overline{OK}}, \\
\ket{\Psi_{OK}}_{\overline{fail}} & =\ket{OK}\ket{W:OK}\ket{\overline{fail}} \ket{\overline{W}: \overline{fail}},   \\
\ket{\Psi_{fail}}_{\overline{fail}} & =\ket{fail}\ket{W:fail} \ket{\overline{fail}}\ket{\overline{W}: \overline{fail}},    
\end{align}
or equivalently expressed in $\overline{F}$ and $F$ basis
\begin{align}
\ket{\Psi_{OK}}_{\overline{OK}} & =\frac{1}{2}\Big[\ket{h}-\ket{t}\Big] \Big[\ket{\downarrow}-\ket{\uparrow}\Big], \\
\ket{\Psi_{fail}}_{\overline{OK}} & =\frac{1}{2}\Big[\ket{h}+\ket{t}\Big] \Big[\ket{\uparrow}-\ket{\downarrow}\Big], \\
\ket{\Psi_{OK}}_{\overline{fail}} & =\frac{1}{2}\Big[\ket{h}-\ket{t}\Big] \Big[\ket{\downarrow}+\ket{\uparrow}\Big], \\
\ket{\Psi_{fail}}_{\overline{fail}} & =\frac{1}{2}\Big[\ket{h}+\ket{t}\Big] \Big[\ket{\downarrow}+\ket{\uparrow}\Big].
\end{align}
Again, since every state is a product state, statements [2.B] and [2.C] cannot be true and the contradiction disappears.
Therefore this way to define agents and to write states after measurements, also erases the contradiction. 

%%%%%%%%%%%%%%%%%%%%

\bigskip

%%%%%%%%%%%%%%%%%%%%

\end{document}